\documentclass[a4, 10pt]{article}

\usepackage{amsmath, amssymb, bm}
\usepackage[top=1in, bottom=1.2in, left=0.5in, right=0.5in]{geometry}
\usepackage[]{graphicx}
\usepackage[caption=false]{subfig}
\usepackage{textcomp}
\usepackage{authblk}
\usepackage{hyperref}
\DeclareMathOperator\arcsinh{arcsinh}

\newcommand{\bs}{\boldsymbol}

\begin{document}

\title{\Large{\bf{Beyond the standard model with sum rules}}}
\author{Damian Ejlli}

\affil{\emph{\normalsize{Department of Physics, Novosibirsk State University, Novosibirsk 630090 Russia}}}

\date{}

\maketitle

\begin{abstract}

The possibility of physics beyond the standard model is studied. The requirement of finiteness of the zero point energy density and pressure or the requirement of the Lorentz invariance of the zero point stress-energy tensor in Minkowski space-time, implies regularization sum rules on the number of degrees of freedom and mass of fundamental particles spectrum. The consequences of these sum rules on the existence of particles beyond the standard model is studied. If these sum rules are to be satisfied, it is shown that some simple and minimal extensions of the standard model such as the two Higgs doublet model, right handed neutrinos, mirror symmetry can not be complete extensions in their current forms. The only exception is unbroken supersymmetry and maybe broken supersymmetry. A comparison between different regularization schemes is also done. It is shown that while all considered regularization schemes give a Lorentz invariant expression for the zero point stress-energy tensor in Minkowski space, their physical interpretations are quite different.

\end{abstract}

\vspace{+1cm}

\section{Introduction}
\label{sec:1}

One of the most important questions of contemporary physics is if does exist an extension of the standard model (SM) of particle physics \cite{Csaki:2016kln}. This question has been so much present for decades and it seems still unclear if such an extension is necessary or not. Indeed, the discovery of a Higgs-like boson particle \cite{Aad:2015zhl} has completed the list of particles that we have been looking for. However, even though the check list of particles seems to be complete, there are some motivations for physics beyond the SM, partially aesthetically in nature, including the hierarchy problem, the strong CP problem, quantum triviality and neutrino oscillations. In addition, the not yet understood nature of dark matter and dark energy has been linked to the possibility of physics beyond the SM.

Typically hints for physics beyond the SM arises mostly due to experimentally observed phenomena such as neutrino oscillations and some theoretical motivations such as the hierarchy problem and the inability to describe physics beyond the Planck scale. However, these facts are not directly a byproduct of the SM itself and it would be desirable if quantum field theory hints for physics beyond the SM. Therefore, is legitimate to ask; is there any particular requirement within quantum field theory that hints for physics beyond the SM? And if yes, are there any current extensions of the SM compatible with such a requirement?

Usually important requirements within a given theory are intrinsically related to the symmetries present within the theory. One particular requirement in quantum field theory, that requires implicitly a symmetry, was suggested several years ago by Pauli \cite{Pauli1951}, who noted that the requirement of the cancellation of the \emph{net} zero point energy density between \emph{fundamental} bosons and fermions, implies sum rules involving the particle masses, their spins and degeneracy factors. Pauli's original idea, was that for each fermion particle must exist a corresponding boson particle, in such a way that the sum of their zero point energy densities must vanish. This cancellation of the net zero point energy density obviously requires an extreme fine tuning between the parameters of the theory. In addition, Pauli's observation was based only in the case of free (non interacting) zero point fields and more importantly the cancellation of the zero point energy density is an \emph{ad hoc} assumption which requires serious motivations.

An important hint that the net zero point energy density might not be exactly zero is related to the fact that according to the theory of general relativity, every form of energy must gravitate, namely it creates spacetime curvature. Based on this fact and on the observational fact that the effective cosmological constant, $\Lambda_\text{eff}$, is non zero, Zeldovich \cite{Zeldovich:1967gd} was one of the first to link the zero point energy density with the cosmological constant. According to his observation, the small observed value of the cosmological constant might be a consequence of a not exact cancellation between the energy densities of fermions and bosons, where the net zero point energy density must be a very small quantity and not necessarily equal to zero. However, this observation is based on the presupposition that $\Lambda_\text{eff}$ is entirely due to the zero point energy density and consequently it could be very well that the zero point energy density could be effectively zero (an extreme fine tuning between fermions and bosons) and $\Lambda_\text{eff}$ gets contribution only from the geometry of spacetime.

The observation made by Pauli \cite{Pauli1951} and the suggestion made by Zeldovich \cite{Zeldovich:1967gd}, tell us an important information, namely that the zero point energy density must be a \emph{finite} quantity. The finiteness of the zero point energy density is an important concept that can be shown to be correct by requiring the Lorentz invariance of the zero point stress energy tensor, $T_V^{\alpha\beta}$, in Minkowski spacetime and vice-versa. Indeed, was still Zeldovich, who showed by using the Pauli-Villars regularization scheme \cite{Zeldovich:1968ehl}, that the net zero point (or vacuum) energy density $\rho_V$ and pressure $P_V$ of free fields satisfy the relation $\rho_V=-P_V$, where $\rho_V$ is a finite quantity different from zero. This fact is of vital importance since one does not necessarily need to have cancellation of $\rho_V$ in order to have a Lorentz invariant vacuum, as one might expect based on simple arguments.

The requirement of the Lorentz invariance of the vacuum, which is a logical requirement for a relativistic invariant  vacuum state, and its possible consequences in particle physics and cosmology, has received a renewed interest in the last decades \cite{Akhmedov:2002ts}-\cite{Koksma:2011cq}. Such a requirement, however, often is presented with serious difficulties which mostly arise due to the regularization schemes used in order to get rid of the ultraviolet divergencies. Indeed, it has been argued in Refs. \cite{Akhmedov:2002ts} and \cite{Ossola:2003ku} that in the cut-off regularization scheme and in the Pauli-Villars regularization scheme, $\rho_V$, diverges quadratically with the renormalization scale while by using the dimensional regularization scheme, $\rho_V$ is logarithmically dependent on the renormalization scale.  On the other hand, in Ref. \cite{Koksma:2011cq} it has been argued that independently on the regularization scheme used, $\rho_V$ is logarithmically dependent on the renormalization scale. In addition, the results found in Refs. \cite{Akhmedov:2002ts} and \cite{Ossola:2003ku} in the case of the Pauli-Villars regularization scheme, seem to contradict the result found by Zeldovich \cite{Zeldovich:1968ehl} which uses the (same) Pauli-Villars regularization scheme and finds a logarithmic dependence on the renormalization scale.

As already mentioned above, Pauli's requirement on the cancellation of $\rho_V$ for free fields, implies sum rules involving the particle masses, spins and degeneracy factors. These sum rules have been taken into consideration by various authors and have been applied for some specific toy models and their possible extension in de Sitter spacetime \cite{Kamenshchik:2006nm} has been studied. However, as pointed out by Visser \cite{Visser:2016mtr} based on Pauli's \cite{Pauli1951} argument, these sum rules can be obtained not necessarily by requiring that $\rho_V=0$ but by requiring the Lorentz invariance of the zero point stress-energy tensor in Minkowski space-time \cite{Visser:2016mtr} or equivalently by requiring the finiteness of $\rho_V$ and $P_V$. The sum rules found by Pauli in case of free fields and extended by Visser to the case of interacting fields, are in reality a cancellation scheme with the final goal to eliminate divergencies and that helps to get rid of the quartic and quadratic divergencies in the cut-off scale. These sum rules, which we are going to derive in the next section, even though are quite simple to treat with, their implications in physics could be of great importance. In fact, as briefly discussed in Ref. \cite{Visser:2016mtr} without any explicit example, these sum rules applied to fundamental particles, hint for physics beyond the standard model.

The sum rules found Pauli and Visser or simply the PV sum rules, define a mutual cancellation scheme in quantum field theory, that regularize divergent integrals and which in principle do not necessarily have to hold. One common assumption in quantum field theory is that independently of the regularization scheme used, the predictions of the theory are the same. However, this assumption is quite ad hoc and there is no a priori way to tell which regularization scheme to use and if at the end they imply the same predictions of the theory, see for example sec. 7.5 of Ref. \cite{Peskin:1995ev} for a discussion on these issues. As we will see in this work the predictions of quantum field theory by using the PV cancellation (regularization) scheme are different from the Pauli-Villars and dimensional regularization schemes. The main goal of this work is to \emph{critically} analyze the PV sum rules and answer to the following questions; are the PV sum rules a consistent regularization scheme to be used in quantum field theory? Assuming that the PV sum rules hold, which are their implications in physics and especially on the physics beyond the standard model? 

Consequently, in this work I consider the possibility of physics beyond the SM based on the assumptions that the zero point stress-energy tensor is Lorentz invariant or that the zero-point energy density and pressure are both finite. These requirements applied to relativistic field theories in Minkowski space-time, imply that particles beyond those included in the SM must exist, if, the PV sum rules hold. After, I study some current extensions of the SM and their compatibility with the PV sum rules. This paper is organized as follows; In Sec. \ref{sec:2}, I discuss that the requirement of the Lorentz invariance of the zero point stress-energy tensor together with the PV sum rules and their implication on physics beyond the SM. In Sec. \ref{sec:3}, \ref{sec:4}, \ref{sec:5} and \ref{sec:6}, I study if the two Higgs doublet model, right handed neutrinos, mirror symmetry and supersymmetry models are compatible with the PV sum rules. In Sec. \ref{sec:7}, I study other regularization schemes other than the PV regularization scheme, and derive regularized expressions for the zero point stress-energy tensor. In Sec. \ref{sec:8}, I conclude. In this paper, I work with the natural units $\hbar=c=k_B=1$ and metric with signature $\eta_{\mu\nu}=\text{diag}(1, -1, -1, -1)$.

\section{Zero point stress-energy tensor and beyond the standard model}
\label{sec:2}

In this section we study the requirement of the Lorentz invariance of the zero point stress-energy tensor in Minkowski space and its implication on the existence of particles beyond the SM. In this section we partially follow Ref.  \cite{Visser:2016mtr}. We start with the expression of the covariant zero point stress-energy tensor, for free fields, in Minkowski space-time which is given by\footnote{In what follows, if not otherwise specified, we will drop the subscript $V$ attached to vacuum energy density $\rho_V=\rho$, pressure $P_V=P$, and vacuum energy momentum tensor $T_V^{\alpha\beta}=T^{\alpha\beta}$.}
\begin{equation}\label{vac-tens}
T^{\alpha\beta}=\sum_n\left[(-1)^{2S_n}g_n\int \frac{d^3k}{(2\pi)^3\,2\,\omega_n(k)}k_n^\alpha k_n^\beta \right],
\end{equation}
where $k^\alpha=(\omega, \bs k)$ with $\omega=\sqrt{m^2+k^2}$ being the total energy of the particle, $g$ being the degeneracy factor, $S$ being the total spin and $\bs k$ being the wave vector. Our goal is to make the Lorentz covariant $T^{\mu\nu}$ in \eqref{vac-tens} also Lorentz invariant. This can be achieved  by first requiring that $T^{\alpha\beta}$ is rotationally invariant, in which case one get $T^{00}=\rho, T^{0\mu}=0, T^{ij}=P\delta^{ij}$, where $\rho$ and $P$ are respectively the \emph{total} energy density and pressure of the zero point system of particles. Their expression are respectively given by
\begin{equation}\label{ene-press}
\rho = \sum_n\left[(-1)^{2S_n}g_n\int \frac{d^3k}{2(2\pi)^3}\sqrt{m_n^2+k^2}\right],\quad P = \sum_n\left[(-1)^{2S_n}g_n\int \frac{d^3k\,k^2}{6(2\pi)^3\sqrt{m_n^2+k^2}}\right].
\end{equation}

In order to evaluate the integrals in \eqref{ene-press}, which are formally divergent, we introduce a cut-off scale, $K$, on the momenta $k$ and then performing the integrals, we get for the sum $\rho+P$
\begin{equation}\label{con-1}
\rho+P = \sum_n\left[ \frac{\pi}{12(2\pi)^3}(-1)^{2S_n}g_n \left(8K^4+4m_n^2 K^2-m_n^4 + O(K^{-2})\right)\right].
\end{equation}
If we require that $T^{\mu\nu}$ must be Lorentz invariant then we must have that $\rho+P=0$. The only possibility for Eq. \eqref{con-1} to be satisfied, is to impose the following on-shell Pauli-Visser (PV) sum rules \cite{Pauli1951} and \cite{Visser:2016mtr} or the PV regularization scheme 
\begin{equation}\label{sum-rules}
\sum_n(-1)^{2S_n}g_n =0, \quad \sum_n(-1)^{2S_n}g_n\, m_n^2=0,\quad \sum_n(-1)^{2S_n}g_n\, m_n^4 = 0.
\end{equation}
There are few important points to discuss at this stage. First, the sum rules in \eqref{sum-rules} are very similar to the constraints founds on the constants and masses of \emph{ghost} particles in the Pauli-Villars regularization scheme, see Sec. \ref{sec:7} for details. In our case the sum rules in \eqref{sum-rules} are extended to \emph{real} elementary particles. Second, the requirement of the Lorentz invariance of $T^{\mu\nu}$ or equivalently $\rho+P=0$, implies that $\rho$ and $P$ must be finite. Indeed, if $\rho$ and $P$ would have to be infinite, we would have an undetermined form, namely $\rho+P=\infty - \infty \neq 0$. Consequently, the Lorentz invariance of $T^{\mu\nu}$ implies finiteness of $\rho$ and $P$. One can also show \cite{Visser:2016mtr} that finiteness of $\rho$ and $P$, imply that $T^{\mu\nu}$ must be Lorentz invariant. 

The key point of the polynomial in mass Eqs. \eqref{sum-rules}, is the sum over the entire spectrum of elementary particles, where \emph{all} sum rules in Eqs. \eqref{sum-rules} must be satisfied simultaneously. The PV sum rules \eqref{sum-rules} have been obtained in the case when particles are on-shell, namely free fields. If we also include field interactions and use renormalization group equation, the PV sum rules \eqref{sum-rules} get modified to \cite{Visser:2016mtr}
\begin{equation}\label{sum-rules-1}
\sum_n(-1)^{2S_n}g_n=0, \quad \sum_n(-1)^{2S_n}g_n\, m_n^2\,\gamma_n=0,\quad \sum_n(-1)^{2S_n}g_n\, m_n^4\,\gamma_n=0,
\end{equation}
where $\mu$ is the renormalization energy scale and the dimensionless $\gamma_n$ functions are the anomalous dimensions defined as usual
\begin{equation}\nonumber
\gamma_n \equiv \frac{\partial \ln m_n}{\partial \ln \mu}=\frac{\mu}{m_n}\frac{\partial m_n}{\partial \mu}.
\end{equation}
We may note that the first sum rule in Eqs. \eqref{sum-rules} does not change under the renormalization group equation and Eqs. \eqref{sum-rules-1} define a class of quantum field theories compatible with the off-shell PV sum rules. It is important to stress that the masses $m_n$ appearing in the sum rules \eqref{sum-rules} are the physical particle masses or simply the pole masses of the free theory. In the sum rules \eqref{sum-rules-1}, $m_n(\mu)$, are the running (renormalizable) particle masses which may coincide with the physical or pole masses depending on the renormalization scheme used, see Ref. \cite{Schwartz:2013pla}.

Having obtained the PV sum rules \eqref{sum-rules}-\eqref{sum-rules-1}, let us now see the implications that have on the physics beyond the SM. Consider the current SM with gauge group $SU(3)\times SU(2)\times U(1)$ composed with three families of leptons $(e, \nu_e), (\mu, \nu_\mu), (\tau, \nu_\tau)$, three families of quarks $(u, d), (c, s), (t, b)$, gauge bosons $g, \gamma, Z^0, W^{\pm}$ and Higgs scalar $H$. In the sum rules\eqref{sum-rules}-\eqref{sum-rules-1} enter the total spin of the particle $S_n$, the mass $m_n$, the degeneracy factor $g_n$ and $\gamma_n$ if particles are off-shell. Let us focus for the moment in the case when particles are on-shell, namely free fields, in which case the sum rules \eqref{sum-rules} would apply. The degeneracy factor includes a spin factor of $g=2S+1$ for massive particles or $g=2$ for massless particles, it also includes a factor of 2 for antiparticles and a factor of 3 due to colour. So the degeneracy factor for photons and gluons is $g=2$, for quark-antiqark is $g=12$, for lepton-antilepton $g=4$, for the $W$ boson $g=6$, for the $Z^0$ boson $g=3$ and Higgs boson $g=1$. 

Regarding the particle physical mass $m_n$, we may note that in the second and third sum rules in \eqref{sum-rules}, particles with the highest mass values give the biggest contributions to the sum rules. Therefore, it is sufficient to consider in the second and third sum rules in \eqref{sum-rules} only the masses \cite{Patrignani:2016xqp} of, $m_c\simeq 1.27$ GeV for the charm quark, $m_b\simeq 4.18$ GeV for the bottom quark, $m_t\simeq 172.44$ GeV for the top quark, $m_\tau\simeq 1.77$ GeV for the $\tau$ particle, $m_{W^{\pm}}\simeq 80.4$ GeV for the $W$ boson, $m_{Z^0}\simeq 91.2$ GeV for the $Z^{0}$ boson and $m_H\simeq 125$ GeV for the Higgs boson. By using this data in the sum rules \eqref{sum-rules}, we get for the elementary particles of the SM only
\begin{eqnarray}\label{SM-sum-rules}
\sum_n^\text{SM}(-1)^{2S_n}g_n &= & -68\neq 0, \nonumber\\  \sum_n^\text{SM}(-1)^{2S_n}g_n\, m_n^2 &\simeq& -277684\; (\text{GeV}^2)\neq 0, \nonumber\\
 \sum_n^\text{SM}(-1)^{2S_n}g_n\, m_n^4 & \simeq &  -9.9\times 10^9\; (\text{GeV}^4)\neq 0.
\end{eqnarray}

As we can see from the expression \eqref{SM-sum-rules}, the (free field) on-shell PV sum rules applied to the SM particles are not satisfied. This fact tells us that there must be more particles, namely the SM particle content is not sufficient to get exact cancellations of the terms in Eqs. \eqref{sum-rules}. Consequently, if the on-shell sum rules \eqref{sum-rules} need to be satisfied, the SM model is not complete and therefore physics beyond it is necessary. Another important fact is that even in the case when particles are off-shell, the sum rule involving only the degeneracy factor is the same as that for on-shell particles, namely the first sum rules in \eqref{sum-rules} and \eqref{sum-rules-1} are identical. This fact is very important because since all sum rules must be satisfied simultaneously and if a model does not satisfy all sum rules in \eqref{sum-rules} and \eqref{sum-rules-1}, it would be incompatible with the PV sum rules.

So far we have seen that the SM does not satisfy all on-shell and off-shell PV sum rules and consequently physics beyond it is necessary. At this point the following question comes naturally; is there any current extension of the SM compatible with the on-shell and off-shell PV sum rules? In order to answer to this question, let us assume that extensions beyond the SM exist. In this case we can divide for example the on-shell sum rules \eqref{sum-rules} into the contribution from the SM particles and those from beyond the SM. Therefore from \eqref{sum-rules} we get
\begin{eqnarray}\label{BSM-sum-rules}
\sum_n^\text{BSM}(-1)^{2S_n}g_n &= & 68, \\ \nonumber \sum_n^\text{BSM}(-1)^{2S_n}g_n\, m_n^2 & \simeq & 277684\; (\text{GeV}^2), \nonumber\\
 \sum_n^\text{BSM}(-1)^{2S_n}g_n\, m_n^4 & \simeq &  9.9\times 10^9\; (\text{GeV}^4)\nonumber,
\end{eqnarray}
where the sum in $n$ in \eqref{BSM-sum-rules} is for those particles not belonging to the SM, namely is a sum for beyond SM particles. 

One important thing which we must be aware before continuing and which we have mentioned above can be stated in the following way: \emph{Based on the PV sum rules \eqref{sum-rules}-\eqref{sum-rules-1}, if a quantum field theory does not satisfy at least the first of free field sum rules in \eqref{sum-rules} (degeneracy factor sum rule), it automatically does not satisfy all the interacting fields sum rules in \eqref{sum-rules-1}}. Based on this fact, in what follows for simplicity in our analysis we consider only the free field sum rule given \eqref{sum-rules} and apply it to beyond SM models.

\section{The Two Higgs doublet model (2HDM)}
\label{sec:3}

Now we consider as possible extension of the SM, the two Higgs doublet model (2HDM), which is the simplest possible  extension of the SM. In this model \cite{Branco:2011iw}, a second Higgs doublet is added to the SM where the most general form of the renormalizable Higgs scalar potential is given by
\begin{eqnarray}\nonumber
V &= & m_{11}^2\Phi_1^\dagger\Phi_1+m_{22}^2\Phi_2^\dagger\Phi_2-m_{12}^2\left(\Phi_1^\dagger\Phi_2+\Phi_2^\dagger\Phi_1\right)\nonumber\\ &+& \frac{\lambda_1}{2}\left(\Phi_1^\dagger\Phi_1\right)^2+\frac{\lambda_2}{2}\left(\Phi_2^\dagger\Phi_2\right)^2+ \lambda_3\Phi_1^\dagger\Phi_1\Phi_2^\dagger\Phi_2\nonumber\\ &+& \lambda_4\Phi_1^\dagger\Phi_2\Phi_2^\dagger\Phi_1+\frac{\lambda_5}{2}\left[\left(\Phi_1^\dagger\Phi_2\right)^2+\left(\Phi_2^\dagger\Phi_1\right)^2 \right]\nonumber,
\end{eqnarray}
where $\Phi_1, \Phi_2$ are two Higgs doublets, $m_{11}, m_{22}, \lambda_1, \lambda_2, \lambda_3$ and $\lambda_4$ are real parameters while the parameters $m_{12}, \lambda_5$ can be either real or complex. The minimization of the potential gives the following expressions for the vacuum expectations values of the doublets $\Phi_1$ and $\Phi_2$
\begin{equation}\nonumber
\langle \Phi_1\rangle =\frac{1}{\sqrt{2}}\begin{pmatrix} 0\\
v_1\end{pmatrix}, \qquad \langle\Phi_2\rangle =\frac{1}{\sqrt{2}}\begin{pmatrix} 0\\
v_2\end{pmatrix},
\end{equation}
where $v^2=|v_1|^2+|v_2|^2=(246$ GeV)$^2$. 

With the introduction of another Higgs doublet, now there are eight fields in total, four for each doublet
\begin{equation}\nonumber
\langle \Phi_k\rangle =\frac{1}{\sqrt{2}}\begin{pmatrix} \sqrt{2}\phi_k^+\\
v_k+\rho_k+i\eta_k\end{pmatrix}, \quad k=1, 2.
\end{equation}
After symmetry breaking, three of these fields get absorbed to form the $W^{\pm}, Z^0$ gauge bosons and there are left only five physical fields at the end. Of these five physical fields, two of them are charged scalar Higgs pair $H^{\pm}$ and the remaining three are neutral scalars $h_1, h_2, h_3$. However, the 2HDM is different with respect to the SM Higgs sector because it allows for Higgs mediated CP violation. If the CP symmetry is not violated\footnote{This assumption is quite common in the 2HDM and in this work we adhere to it. Usually is also assumed that CP is not spontaneously broken.}, the three neutral scalars $h_1, h_2, h_3$ can be classified as two CP even scalars $h_1=H, h_2=h$ and one CP odd scalar $h_3=A$ with the mass condition (by convention) $m_H>m_h$. So, at the end in the 2HDM we have two charged scalars $H^{\pm}$ with masses $m_{H^{\pm}}$, two neutral scalars $h, H$ with masses $m_h, m_H$ and a neutral pseudoscalar $A$ with mass $m_A$. In addition one of the two neutral scalars $h$ or $H$ is identified with the observed Higgs-like boson which we choose to be $m_H$ with mass $m_H\simeq 125$ GeV. 

At this point let us ask the question; does the 2HDM satisfy the on-shell and off-shell PV sum rules? In order to answer to this question, let us consider that the neutral Higgs $H$ with mass $m_H$ belongs to the SM and the remaining four Higgs particles belong to beyond the SM. In this case, for the Higgs particles $H^{\pm}, h, A$ belonging to beyond the SM, the on-shell PV sum rules \eqref{BSM-sum-rules} give
\begin{eqnarray}\label{2HDM}
\sum_n^\text{BSM}(-1)^{2S_n}g_n &= & g_{H^{\pm}}+g_h+g_A = 4 \neq	68, \\
 \sum_n^\text{BSM}(-1)^{2S_n}g_n\, m_n^2 &= & 2 m_{H^{\pm}}^2+m_h^2+m_A^2 \simeq 277684\; (\text{GeV}^2),\nonumber\\
 \sum_n^\text{BSM}(-1)^{2S_n}g_n\, m_n^4 &= & 2 m_{H^{\pm}}^4+m_h^4+m_A^4 \simeq  9.9\times 10^9\; (\text{GeV}^4),\nonumber
\end{eqnarray}
where $g_{H^\pm}=2$, $g_h=1$ and $g_A=1$ since all Higgs particles $H^{\pm}, h, A$ are scalar bosons with spin $S=0$. The value of $g_{H^\pm}=2$ takes into account that $H^\pm$ is a charged scalar field where $H^+$ is the antiparticle of $H^-$ and vice versa. As we can see from \eqref{2HDM}, the first on-shell PV sum rule is not satisfied by the 2HDM. Since the first sum rule is the same also for off-shell particles, we can conclude that the 2HDM does not satisfy all on-shell and off-shell PV sum rules. The second and third sum rules in \eqref{2HDM} can be seen as equations for the unknown masses $m_{H^\pm}, m_h$ and $m_A$. However, since there are two equations for three unknown variables, the system is undetermined. If one knows the mass of any of the Higgs scalars $m_{H^\pm}, m_h, m_A$, the system of equations can be reduced to two unknown parameters.

\section{Right handed neutrinos}
\label{sec:4}

Another minimal and simple extension of the SM is to add right handed neutrinos in order to explain the observed phenomena of neutrino oscillations. The fact that known neutrinos oscillate means that they have a mass that usually is very small, namely of the order of eV or even smaller. The best known renormalizable extension of the SM which generates small neutrino masses is the so called seesaw (type I and type II) mechanism\cite{Yanagida:1980xy}. Without going into details of how the seesaw mechanisms works, for our purposes is important the number of right handed neutrinos added to the SM, their masses, spins and degeneracy factors. The minimum number of right handed neutrinos in order to generate the masses of known left handed neutrinos of the SM is of two. 

Assume for simplicity that there are three right handed neutrinos, one for each left handed neutrino and let $m_{R 1}, m_{R 2}$ and $m_{R 3}$ be their masses. In addition, the right handed neutrinos have to be sterile in order to not participate in the SM interactions, namely they must have no charge and be their own antiparticles.
Since right handed neutrinos are fermions with spin $S=1/2$, massive and with no charge, their degeneracy factors are all equal to $g=2S+1=2$. With three right handed neutrinos, the on-shell PV sum rules \eqref{BSM-sum-rules} read
\begin{eqnarray}\label{sterile}
\sum_n^\text{BSM}(-1)^{2S_n}g_n &= & -g_{R1}-g_{R2}-g_{R3} = -6 \neq	68, \\
 \sum_n^\text{BSM}(-1)^{2S_n}g_n\, m_n^2 &= & -2 m_{R1}^2-2m_{R2}^2-2m_{R3}^2 \simeq 277684\; (\text{GeV}^2),\nonumber\\
 \sum_n^\text{BSM}(-1)^{2S_n}g_n\, m_n^4 &= & -2 m_{R1}^4-2m_{R2}^4-2m_{R3}^4 \simeq  9.9\times 10^9\; (\text{GeV}^4).\nonumber
\end{eqnarray}

We may observe from \eqref{sterile} that none of the on-shell PV sum rules is satisfied with the extension of the SM by introducing right handed neutrinos. The first sum rule in \eqref{sterile} for beyond the SM particles has opposite sign with respect to the part including SM particles. Moreover, the second and third sum rules in \eqref{sterile} have no solutions in terms of right handed neutrino masses since the sum of their masses to power two and four must be negative. Obviously, one would expect this behaviour since right handed neutrinos are fermions and consequently they account for negative signs in the PV sum rules. Even if we consider just two or more than three right handed neutrinos, we would arrive to the same conclusions, namely none of the on-shell PV sum rules are satisfied. In the case of off-shell PV sum rules, the first sum rule is still not satisfied while the other sum rules might be satisfied or not depending on the signs of $\gamma_n$ and their respective values.

\section{Mirror symmetry}
\label{sec:5}

The concept of mirror symmetry or more precisely parity symmetry relies on the fact that if parity, namely $\bs x\rightarrow -\bs x$, is a symmetry of the total Lagrangian, then new physics must exist and a new sector of particles, the so called mirror sector is invoked \cite{Foot:1991bp}. In general the Lagrangian of the SM is not invariant under parity symmetry and consequently parity is violated in the SM. In order to have a model where parity is not violated, one adds to the SM Lagrangian, $\mathcal L_1$, a new Lagrangian $\mathcal L_2$ and an interaction term $\mathcal L_\text{int}$. In this case the total Lagrangian of the model is $\mathcal L=\mathcal L_1+\mathcal L_2+\mathcal L_\text{int}$ and it is parity invariant provided that $\mathcal L_1\leftrightarrow \mathcal L_2$ under parity symmetry and if $\mathcal L_\text{int}$ contains mixing terms that are parity invariant.
 
More formally, the mirror symmetry model assumes that there exist a larger symmetry group which is given by the product of $G\times G^\prime$, where $G=SU(3)\times SU(2)\times U(1)$ is the gauge group of the SM and $G^\prime=SU^\prime(3)\times SU^\prime(2)\times U^\prime(1)$ is the gauge group of the mirror sector. Under parity symmetry $P$, $P(G\leftrightarrow G^\prime)$. Being parity $P$ a symmetry of $G\times G^\prime$, means that both sectors are described by the same Lagrangians and all coupling constants of the theory have the same pattern. The parity operation on $\mathcal L_1$, transforms left handed (right handed) fermions into right handed (left handed) fermions and vice versa for $\mathcal L_2$. More precisely under $\bs x\rightarrow -\bs x$ and $t\rightarrow t$, we have for fermions $f_L\leftrightarrow \gamma_0 f_R^\prime, q_L\leftrightarrow \gamma_0 q_R^\prime, u_R\leftrightarrow \gamma_0 u_L^\prime, e_R\leftrightarrow \gamma_0 e_L^\prime, d_R\leftrightarrow \gamma_0 d_L^\prime$ and  $G^\mu\leftrightarrow G^{\prime\mu}, W^\mu\leftrightarrow W^{\prime\mu}, B^\mu\leftrightarrow B^{\prime\mu}$ for gauge bosons of the theory. In addition, the model has two Higgs doublets, one for each sector $\phi\sim (1, 2, 1)$ and $\phi^\prime\sim (1, 2, -1)$.

The $\bs Z_2$ symmetry or parity symmetry of the product group $G\times G^\prime$ can be exact or spontaneously broken. If parity is an exact symmetry, the particles of the mirror sector have the same masses, interactions and coupling constants as their corresponding SM particles. For our purposes, we need only the masses spins and degeneracy factors of the mirror sector particles. The fact that mirror sector particles have the same masses, spins and same degeneracy factors for unbroken parity symmetry, means that none of the on-shell PV sum rules are satisfied with the introduction of the mirror sector, namely
\begin{eqnarray}\label{mirror-sum-rules}
\sum_n^\text{BSM}(-1)^{2S_n}g_n &= & -68\neq 68,\\  \sum_n^\text{BSM}(-1)^{2S_n}g_n\, m_n^2 &\simeq & - 277684\; (\text{GeV}^2)\neq 277684 \; (\text{GeV}^2), \nonumber\\
 \sum_n^\text{BSM}(-1)^{2S_n}g_n\, m_n^4 & \simeq &  -9.9\times 10^9\; (\text{GeV}^4) \neq 9.9\times 10^9\; (\text{GeV}^4)\nonumber,
\end{eqnarray}
where in the sum in \eqref{mirror-sum-rules} we considered the mirror sector particles as beyond the SM particles.  In the case of off-shell particles and unbroken parity symmetry, the first sum rule over the degeneracy factor is identical to the first sum rule in \eqref{mirror-sum-rules}  which is not satisfied, while the second and third sum rules might be satisfied or not depending on the signs of $\gamma_n$.
 
 If parity symmetry is broken at some energy scale $\Lambda$, then particle masses of the mirror sector need not to be equal to the SM particle masses. However, since the mirror sector is a copy of the SM sector, the degeneracy factors and spins of particles of the mirror sector are exactly the same as those of the SM particles even after mirror symmetry breaking. This means that for broken parity symmetry, the first PV sum rule is not satisfied for both on-shell and off-shell particles of the mirror sector. On the other hand, the second and third PV sum rules might be satisfied or not for on-shell and off-shell mirror sector particles. 

Above we treated the simplest realization of mirror symmetry where two Higgs doublets are present in total. Another possible realization of mirror symmetry is the so called Twin Higgs Model \cite{Barbieri:2016zxn}. One can check that even this realization of mirror symmetry does not satisfy none of on-shell PV sum rules. For off-shell particles the first sum rule is not satisfied while the second and the third might be satisfied or not.

\section{Supersymmetry}
\label{sec:6}

Probably one of the most important extensions of the standard model is supersymmetry. In the theory of supersymmetry every SM particle has a superpartner\footnote{Also each SM antiparticle has its corresponding supersymmetric antiparticle. } that has opposite statistics and different spin, for a review see Ref. \cite{Martin:1997ns}. In the case of unbroken supersymmetry, all superpartner particles have the same masses and the same internal quantum numbers except the spin. In the context of PV sum rules, the most important quantities for us are the spin of supersymmetric particles their degeneracy factors and masses. Indeed, for all fermions (quarks and leptons) which have spin 1/2, their corresponding superpartner  boson particles, squarks and sleptons have spins that differ by half integer, namely sleptons and squarks all have spin equal to zero. This fact is of fundamental importance since the degeneracy factors corresponding to the particle spin part does cancel. 

In order to see more closely what has been stated above in the case of unbroken supersymmetry, consider for example the case of leptons that all have degeneracy factors, $g_{leptons}=2(2S+1)=4$, and their corresponding superpartners, namely sleptons, all having degeneracy factors $g_{sleptons}=2\times 2(2S+1)=4$. The factor of two in front of $2S+1$ in case of fermions takes into account the antiparticle contribution and the the additional factor of two in case of corresponding superpartners comes from the fact that there are two scalar supersymmetric particles for each fermion. So, in the case of leptons and sleptons, the first on-shell PV sum rule would give, $\sum_{n}^{leptons}(-1)^{2S_n} g_n+\sum_{n}^{sleptons}(-1)^{2S_n} g_n=-6\times 4+6\times 2\times 2= 0$, where the factor of six takes into account that there are six leptons in the SM and twelve sleptons for beyond the SM. One can proceed in the same way for the remaining particles and conclude that in the case of unbroken supersymmetry, the first on-shell PV sum rule is satisfied. Moreover, since $g_n$ appears also in the second and third PV sum rules and being the masses of supersymmetric particles equal to those of the SM in case of unbroken supersymmetry, implies also that the second and third on-shell PV sum rules are satisfied. Thus, the on-shell PV sum rules are satisfied for unbroken supersymmetry. In the case of broken supersymmetry, the first on-shell and off-shell sum rule are automatically satisfied while the other sum rules might be satisfied or not.

\section{Regularization schemes}
\label{sec:7}

In the previous sections we applied the PV sum rules to some possible extensions of the SM and saw that only unbroken and perhaps broken supersymmetry may be compatible with the on-shell and off-shell PV sum rules \eqref{sum-rules} and \eqref{sum-rules-1}. The conclusions that we have found in the previous sections, however, depend on the regularization scheme used to get rid of the quadratic, $K^2$, and quartic, $K^4$, divergencies in \eqref{con-1}, where the PV sum rules are indeed a mutual cancellation scheme of divergencies that can also been seen as a regularization scheme. In the context of the PV sum rules, we can calculate for the zero point energy density of non interacting fields 
\begin{equation}\label{en-densi}
\begin{gathered}
\rho=\sum_n\left[(-1)^{2S_n}g_n\int \frac{d^3k}{2(2\pi)^3}\sqrt{m_n^2+k^2}\right]=\sum_n\left[(-1)^{2S_n}g_n\frac{\pi}{2(2\pi)^3}\left(K^3 \sqrt{K^2+m_n^2}+\frac{1}{2} K m_n^2 \sqrt{K^2+m_n^2}\right.\right. \\ \left.\left. -\frac{1}{2} m_n^4\arcsinh\left(\frac{K}{m_n}\right) \right) \right] \simeq \sum_n\left[(-1)^{2S_n}g_n\frac{\pi}{2(2\pi)^3}\left(K^4 + K^2 m_n^2+\frac{1}{8}m_n^4 + \frac{1}{4}m_n^4\ln\left(\frac{m_n^2}{4 K^2}\right) +O\left(\frac{1}{K^2}\right)\right) \right],
\end{gathered}
\end{equation}
where we used series expansion in the term $m_n/K\ll 1$. If now we use the on-shell PV sum rules \eqref{sum-rules} in the first three terms on the right hand side of expression \eqref{en-densi}, we get for the zero point energy density the final expression
\begin{equation}\label{density}
\rho\simeq \frac{1}{64\pi^2} \sum_n (-1)^{2S_n}g_n\,m_n^4 \ln\left(\frac{m_n^2}{4 K^2}\right).
\end{equation}
We can proceed in the same way as above for the calculation of the zero point pressure, where we obtain
\begin{equation}\label{pressure}
P=\sum_n\left[(-1)^{2S_n}g_n\int \frac{d^3k\,k^2}{6(2\pi)^3\sqrt{m_n^2+k^2}}\right]\simeq - \frac{1}{64\pi^2} \sum_n (-1)^{2S_n}g_n\,m_n^4 \ln\left(\frac{m_n^2}{4 K^2}\right),
\end{equation}
where we still used the on-shell PV sum rules \eqref{sum-rules} in order to get rid of the quadratic and quartic divergencies in $K$.

The expressions for the zero point energy density and pressure calculated in \eqref{density}-\eqref{pressure} imply that these quantities are finite and have opposite signs with respect to each other, namely we get that $\rho=-P$, which is the correct condition for the Lorentz invariance or relativistic invariance of the zero point stress energy tensor, where 
\begin{equation}\label{Pauli-Visser}
T^{\alpha\beta} \simeq \frac{1}{64\pi^2} \sum_n (-1)^{2S_n}g_n\,m_n^4 \ln\left(\frac{m_n^2}{4 K^2}\right)\,\eta^{\alpha\beta} \quad (\text{Pauli-Visser})
\end{equation}
Therefore, the PV sum rules \eqref{sum-rules} are the crucial conditions in order to get the relativistic invariance and finiteness of $\rho$ and $P$. In Refs. \cite{Akhmedov:2002ts}-\cite{Koksma:2011cq} similar expressions for $\rho$ and $P$ as in \eqref{ene-press} are presented and there is no weighted sum over the fundamental particle spectrum. In Refs. \cite{Akhmedov:2002ts}, \cite{Ossola:2003ku} and \cite{Koksma:2011cq} is concluded that the introduction of a three dimensional cut-off moment $K$ would break the relativistic invariance and therefore this scheme is not suitable for regularization unless one introduce ad hoc counter terms. However, as we have showed so far, we obtain a perfectly Lorentz invariant or relativistic invariant result, even in the case when it is introduced a three dimensional cut-off scale as far as the PV sum rules and the weighted sum over all fundamental particles are used. 

In order to see the importance of the PV sum rules and the weighted sum on particle species, let us follow Refs. \cite{Akhmedov:2002ts}, \cite{Ossola:2003ku} and \cite{Koksma:2011cq} and consider for example a single scalar field, $(-1)^{2S}=1$,  which the sum of the zero point energy density and pressure is given by 
\begin{equation}\label{scalar-field}
\rho^{(s)}+P^{(s)}=\frac{1}{12\pi^2}\,K^3\sqrt{m^2+K^2}\neq 0 \quad \text{for}\quad K\rightarrow \infty.
\end{equation}
As we can see from \eqref{scalar-field}, there is no way for a single scalar field to satisfy the Lorentz invariant expression $\rho+P=0$. A Lorentz invariant result, in the case when we use a three dimensional cut-off, $K$, is obtained not by considering a \emph{single} species but using the PV sum rules together with the weighted sum over \emph{all} fundamental particle species.

So far we have discussed about the PV cancellation (regularization) scheme and it is very important to confront it with other well known regularization schemes; the Pauli-Villars and the dimensional regularization schemes. These regularization schemes are based on a covariant formulation and do not make use of three dimensional cut-off scale, $K$.  In order to make the Pauli-Villars and the dimensional regularization schemes more comprehensible, let us write the zero point stress-energy tensor of free fields in \eqref{vac-tens} in four dimensional representation as 
 \begin{equation}\label{vac-tens-0}
T^{\alpha\beta}=\sum_{n=1}^N \left[(-1)^{2S_n}g_n\int \frac{d^4 k}{(2\pi)^3}\,k_n^\alpha k_n^\beta\, \delta (k_n^2-m_n^2)\, \theta(k_n^0)\right],
\end{equation} 
where $\theta(k_n^0)$ is the Heaviside theta function and $\delta(k_n^2-m_n^2)$ is the Dirac delta function with $k_n^2={(k_\mu k^\mu)}_n$. In the Pauli-Villars regularization scheme \cite{Pauli:1949zm}, one introduce a set of \emph{artificial} regulator fields, with masses $M_j$ and with factors $C_j\, (j=1, 2,..., L)$ with the appropriate statistics for bosons and fermions. In this case the regularized zero point stress-energy tensor, $T^{\alpha\beta}$, in \eqref{vac-tens-0} becomes
 \begin{equation}\label{vac-tens-1}
 T^{\alpha\beta}= \sum_{n=1}^N \left\{ \left[(-1)^{2S_n}g_n \int \frac{d^4 k}{(2\pi)^3}\,k_{n}^\alpha k_{n}^\beta\, \delta (k_{n}^2-m_{n}^2)\, \theta(k_{n}^0)\right]+ \sum_{j=1}^{L} \left[ C_{n, j} \int \frac{d^4 k}{(2\pi)^3}\,k_j^\alpha k_j^\beta\, \delta (k_j^2-M_{n, j}^2)\, \theta(k_j^0)\right] \right\},
\end{equation} 
where $L>N$, and the factors $C_j$ satisfy the following conditions:
\begin{equation}\label{PVi-sum-rules}
 \sum_{n=1}^N \left[(-1)^{2S_n} g_n + \sum_{j=1}^L C_{n, j}\right]=0, \quad \sum_{n=1}^N \left[(-1)^{2S_n} g_n m_n^2 + \sum_{j=1}^L C_{n, j} M_{n, j}^2\right]=0, \quad  \sum_{n=1}^N \left[(-1)^{2S_n} g_n m_n^4 + \sum_{j=1}^L C_{n, j} M_{n, j}^4\right]=0.
\end{equation}

We may observe that Eqs. \eqref{PVi-sum-rules} have close resemblance with the on-shell PV sum rules \eqref{sum-rules}, where the sum on $j$ of $C_{n, j}, C_{n, j} M_{n, j}^2, C_{n, j} M_{n, j}^4$ play the role of BSM particles in the case the sum in $n$ is extended to only to the SM particles. We can use the sum rules \eqref{PVi-sum-rules} in \eqref{vac-tens-1} and then by proceeding in the same way as in Ref. \cite{Ossola:2003ku}, we get the following expression for the Pauli-Villars regularized zero point stress-energy tensor
\begin{equation}\label{regu-PVi-tens}
T^{\alpha\beta} = \frac{\eta^{\alpha\beta}}{64\pi^2} \sum_{n=1}^N  \left[ (-1)^{2S_n} g_n m_n^4 \ln\left(\frac{m_{n}^2}{\nu^2}\right)+ \sum_{j=1}^L C_{n, j}\,M_{n, j}^4 \ln\left(\frac{M_{n, j}^2}{\nu^2}\right)\right] \quad \text{for}\quad M_{n, j}\rightarrow \infty, \quad (\text{Pauli-Villars})
\end{equation}
where $\nu$ is a fixed energy/mass scale. The appearance of the fixed parameter $\nu$ in \eqref{regu-PVi-tens} is only formal and it does not contribute to the stress-energy tensor. Indeed, by expanding the logarithms in \eqref{regu-PVi-tens} and using the last sum rule in \eqref{PVi-sum-rules}, the contribution of $\nu$ cancels out. The expression for $T^{\alpha\beta}$ in \eqref{regu-PVi-tens} has a close resemblance with the PV regulated $T^{\alpha\beta}$ in \eqref{Pauli-Visser}, with the main difference that in the Pauli-Villars regularization scheme are introduced artificially regulator fields. Obviously, $T^{\alpha\beta}$ in \eqref{regu-PVi-tens} is a Lorentz invariant quantity and in principle we may add to the sum rules in \eqref{PVi-sum-rules}, an additional condition on the logarithmic term in \eqref{regu-PVi-tens} of the form
\begin{equation}\label{log-cond}
\sum_{n=1}^N \sum_{j=1}^L C_{n, j}\,M_{n, j}^4 \ln\left(\frac{M_{n, j}^2}{\nu^2}\right)=0.
\end{equation}
The additional condition \eqref{log-cond}, explicitly guarantees that in \eqref{regu-PVi-tens} contribute only real particles to the zero point stress-energy tensor. If we consider for example only a scalar field, we would need at least three regulator fields $C_{1, j}, M_{1, j}, j\geq 3$ if we consider only the sum rules in \eqref{PVi-sum-rules} or at least four regulator fields if the additional condition \eqref{log-cond} is taken into account. If there are more than one particles, $n> 1$, the number of regulator fields increases considerably and the Pauli-Villars regularization scheme becomes quaite difficult to handle with.

Another important regularization scheme widely used in quantum field theory is the dimensional regularization \cite{tHooft:1972tcz}. The basic idea of this regularization scheme is to find a spacetime dimension $d$ where a given integral is convergent and then by analytic continuation evaluate the integral at the desired dimension. In order to use the dimensional regularization scheme, let us rewrite \eqref{vac-tens} as
 \begin{equation}\label{vac-tens-2}
T^{\alpha\beta}=\sum_{n=1}^N \left[(-1)^{2S_n}g_n \left(\frac{\eta^{\alpha\beta} m_n^2}{4}\right) \int \frac{d^4 k}{(2\pi)^4}\,\frac{i}{k_n^2-m_n^2+i\epsilon}\right],
\end{equation} 
where we used $k_n^\alpha k_n^\beta=k_n^2 \eta^{\alpha\beta}$ and the property $k_n^2\, \delta(k_n^2-m_n^2)=m_n^2\, \delta (k_n^2-m_n^2)$. The next step is to evaluate the integral on the right hand side of \eqref{vac-tens-2} by making use of Wick rotation, $k^0=ik_E^0, |\bs k|=|\bs k_E|$, namely we can write 
\begin{equation}\nonumber
\lim_{\epsilon\rightarrow 0}\int \frac{d^4 k}{(2\pi)^4}\,\frac{i}{k_n^2-m_n^2+i\epsilon}=\int \frac{d^4 k_E}{(2\pi)^4}\,\frac{1}{k_{E, n}^2+m_n^2}.
\end{equation}
Now we can move to a dimension $d$ and use the well know result 
\begin{equation}\nonumber
\int \frac{d^d k_E}{(2\pi)^d}\,\frac{1}{k_{E, n}^2+m_n^2}=\frac{1}{(4\pi)^{d/2}}\Gamma(1-d/2) m_n^{d-2}.
\end{equation}
By making use of the above results, the expression for the zero point stress-energy tensor in $d$ dimensions becomes
\begin{equation}\label{vac-tens-3}
T^{\alpha\beta}= - \sum_{n=1}^N \left[(-1)^{2S_n}g_n m_n^4 \, (4\pi)^{-d/2} \Gamma(-d/2) (\mu_*/m_n)^{4-d}\right]\left(\frac{\eta^{\alpha\beta}}{2}\right),
\end{equation} 
where $\mu_*$ is an energy scale which has been introduced to take into account the physical dimensions of $T^{\alpha\beta}$ in $d$ dimensions and used the property $\Gamma(1-d/2)=-(d/2)\Gamma(-d/2)$. The next step is to use another property of the Gamma function, namely the recurrence property for negative numbers $\Gamma(z)=\Gamma(1+n+z)/[z(1+z)....(n+z)]$ where $n$ is an integer number and $z$ is a complex number. In our case we can write for $\epsilon\ll 1$ and in the neighbourhood of $-d/2$
\begin{equation}\nonumber
\Gamma(-d/2)=-\frac{8}{d}\frac{\Gamma(3-d/2)}{(2-d)(4-d)}\simeq \frac{1}{\epsilon}\left(\frac{8}{8-6\epsilon+\epsilon^2}\right)\left(\frac{8}{8+6\epsilon+\epsilon^2}\right) \left[ 1+\frac{3\,\epsilon}{4} - \frac{\gamma_\text{M}\epsilon}{2} - \frac{3 \gamma_\text{M} \epsilon^2}{8} -\frac{\gamma_\text{M}^2 \epsilon^2}{4}  \right],
\end{equation}
where we expressed $\epsilon \equiv 4-d$, $\gamma_\text{M} \simeq 0.5772$ is the Euler-Mascheroni constant and used the Weierstraass formula $\Gamma^{-1}(z)=ze^{\gamma_\text{M}z}{\displaystyle \prod_{n=1}^\infty} \left( 1+z/n\right)e^{-z/n}$ expressed in terms of $\epsilon\ll1$. Also for $\epsilon\ll 1$, we can expand in series $(4\pi)^{-d/2}= (4\pi)^{-2+\epsilon/2} \simeq (4\pi)^{-2} \left[1+ (\epsilon/2)\ln(4\pi) +O(\epsilon^2) \right]$ and $(\mu_*/m_n)^\epsilon \simeq 1+\epsilon \ln(\mu_*/m_n) +O(\epsilon^2) $ and then by taking the limit $d\rightarrow 4$ from below or $\epsilon\rightarrow 0^+$, we get for the zero point stress-energy tensor given in \eqref{vac-tens-3} the following expression
\begin{equation}\label{vac-tens-4}
T^{\alpha\beta} \simeq - \sum_{n=1}^N \left[(-1)^{2S_n}g_n m_n^4 \, \left( \frac{2}{\epsilon}+ \frac{3}{2} - \gamma_\text{M} - \ln\left(\frac{m_n^2}{4\pi \mu_*^2}\right) \right) \right]\frac{\eta^{\alpha\beta}}{64 \pi^2} \quad \text{for} \quad \epsilon\rightarrow 0 \quad (\text{dimensional regularization}).
\end{equation} 

We may notice that the expression in \eqref{vac-tens-4} is Lorentz invariant and is not identical to those obtained by using the PV and the Pauli-Villars regularization schemes because of the presence of the pole and of the constant factors of $3/2$ and $\gamma_M$ in \eqref{vac-tens-4}. It is through the renormalization procedure that the pole for $\epsilon\rightarrow 0$ and some constant factors are removed. We must stress that it is after the renormalization procedure that the regularized zero point stress-energy tensors obtained through the Pauli-Villars \eqref{regu-PVi-tens} and dimensional regularization \eqref{vac-tens-4} coincide with the expression of the zero point stress-energy tensor obtained by using the PV cancellation scheme \eqref{Pauli-Visser}. In the expression \eqref{Pauli-Visser} the dependence on $K\rightarrow \infty$ does not contribute to the stress-energy tensor since by expanding the logarithmic term it vanishes because of the third sum rule in \eqref{sum-rules}.

\section{Conclusions}
\label{sec:8}

 In this work we have studied the implications that arises by \emph{assuming} the validity of the PV sum rules \eqref{sum-rules}-\eqref{sum-rules-1}. These sum rules are a consequence of the requirement that the zero point stress-energy tensor in Minkowki spacetime is Lorentz invariant or equivalently the zero point energy density and pressure are finite. These sum rules are polynomial in mass equations that involve the spin, mass and degeneracy factors of all zero point state particles. These sum rules are quite simple, very powerful and allowed us to deduce that particles beyond the SM must exist, if \eqref{sum-rules}-\eqref{sum-rules-1} have to be satisfied. In addition, if \eqref{sum-rules}-\eqref{sum-rules-1} have to hold, the SM particle stress-energy tensor alone is not Lorentz invariant and therefore in order to get a Lorentz invariant zero point stress-energy tensor, beyond SM particles are a necessary condition.

Then we studied several extensions of the SM and if they satisfy the PV sum rules \eqref{sum-rules}-\eqref{sum-rules-1}. We have found that none of the models (except unbroken and maybe broken supersymmetry) satisfy all sum rules simultaneously. Indeed, the first model that we studied, namely the 2HDM, which adds an additional Higgs doublet to the SM, does not satisfy the first on-shell and off-shell PV sum rules while the other two sum rules might be satisfied or not. The second model which we studied is by considering right handed neutrinos as possible extension of the SM. However, such extension is not compatible with PV sum rules and in addition all on-shell sum rules for right handed neutrinos have same (negative) signs with respect to on-shell sum rules for SM particles. The third model that we have studied is the mirror symmetry or parity symmetry. This symmetry has some interesting properties since if it is unbroken, mirror sector particles have the same masses and same internal quantum numbers as SM particles. However, such feature is not enough to have cancellations in the PV sum rules \eqref{sum-rules}-\eqref{sum-rules-1}. Indeed, this arises because mirror sector particles have the same spins and same degeneracy factors as their corresponding SM particles. In the case of supersymmetry, the fact that for fermion particles of the SM correspond two scalar boson particles as beyond the SM and vice versa for boson particles of the SM, means that in the PV sum rules these two families have opposite sign with respect to each other which is a necessary condition for cancellation. This fact does allows exact cancellation for all \eqref{sum-rules}  sum rules if supersymmetry is unbroken. In the case of broken supersymmetry the first rule is satisfied while the others may or not be satisfied.

 One of the key aspects of the PV sum rules is that the first sum rule which involves only the particle number of degrees of freedom survives the renormalization group equation and it remains invariant. Since this particular sum rule is the same in free and interacting quantum field theories, it can be used immediately to exclude those extensions of the SM that are incompatible with it. This fact is the main reason why in analyzing the PV sum rules we concentrated only on the non interacting sum rules \eqref{sum-rules}. This sum rule which essentially gets rid of the quartic divergencies in $\Lambda$ must be satisfied together with the rest of the PV sum rules, if we want to have a Lorentz invariant zero point stress energy tensor (in the context of the PV sum rules). In order to illustrate the importance of this particular sum rule, we may already realize from Eqs. \eqref{2HDM} in the case of the two Higgs doublet model that in principle there might exist masses, $m_{H^\pm}, m_h, m_A$, which may satisfy the second and the third PV sum rules, but there is no way that this model satisfy the first sum rule in either free or interacting field theory.
 
 Whether the first PV sum rule is satisfied or not, intrinsically seems to be related with an underlying symmetry between fermions and bosons. For this reason, supersymmetry is probably the best candidate which may satisfy all three PV sum rules. Indeed, it has already been known since the first formulation of the supersymmetry theory, that the exact cancellation between fermions and bosons helps to get rid of ultraviolet divergencies in loop integrals. In addition, if supersymmetry is unbroken, it has been shown by Zumino \cite{Zumino:1974bg} that the zero point energy density must be zero. The latter fact, immediately can be verified from the PV sum rules, where all sum rules are satisfied for unbroken supersymmetry. In addition, if supersymmetry is unbroken, also the logarithmic in mass term in \eqref{density} must vanish identically, which implies that $\rho=P=0$. However, since unbroken supersymmetry is in direct conflict with experiments, it seems quite plausible that it may be broken at some energy scale and consequently the logarithmic in mass term in \eqref{density} does not necessarily have to be zero as far as supersymmetry is concerned.

 Having discussed the implications on beyond the SM physics by assuming that the PV sum rules are valid, the other main question to answer is if the PV sum rules are a consistent mutual cancellation sceheme in quantum field theory? The answer of this question essentially depends on the particular quantum field theory and if it is finite or not. In fact, it has already been pointed out in Ref. \cite{Visser:2016mtr} that certainly there do exist finite and not finite quantum field theories that satisfy the sum rules \eqref{sum-rules-1}. These field theories are those (supersymmetric) which in \eqref{sum-rules-1}  have either all $\gamma_n=0$ or all $\gamma_n$ equal.

Another important aspect of the PV sum rules is that they have been derived by requiring that the zero point stress-energy tensor is Lorentz invariant, or more precisely, we derived these sum rules in order to make $T^{\alpha\beta}$ Lorentz invariant. However, as we have seen in Sec. \ref{sec:7}, we can get a Lorentz invariant $T^{\alpha\beta}$ without using the PV sum rules, namely by using the Pauli-Villars and the dimensional regularization methods. As we have shown in Sec. \ref{sec:7}, all three methods used give at the end a Lorentz invariant $T^{\alpha\beta}$ but the physical intepretations and consequences of these regularization methods are very different. This fact is very important since in those cases where the on-shell PV sum rules have been used in the literature \cite{Kamenshchik:2006nm}, it has been tacitly assumed that the on-shell PV sum rules are valid. However, as we have shown this is not necessarily true since in the case we adopt conventional regularization methods, the PV sum rules do not arise at all. Moreover, if we adopt the conventional regularization methods as presented in Sec. \ref{sec:7}, all conclusions made so far based on the PV sum rules regarding the physics beyond the SM etc., would be invalid and non existing. 
 
Probably one of the main reasons to adopt the PV sum rules is related to the fact that they have a direct physical interpretation that involve real particles. On the other hand, the Pauli-Villars regularization method involves artificial or ghost fields that have not any apparent physical interpretation. In the case of the dimensional regularization, we move to a fake dimension $d$ where the integrals converge and then by an analytic continuation we move to the physical dimension which is $d=4$. In addition, in the dimensional regularization method one has to artificially add to some degree counter terms in order to get rid of the divergent term and to some constant terms. So, seen in this way, the PV cancellation scheme (or regularization scheme), seems to be more ''physical'' than the Pauli-Villars and dimensional regularization schemes. On the other hand, the dimensional regularization method is known to work also for interacting fields and to all orders in perturbation theory without requiring the existence of extra particles, while the PV cancellation scheme requires the existence of extra particles and hold in the case of interacting fields\footnote{Recently it has been shown in Refs. \cite{Kamenshchik:2018ttr} for the case of some toy models, that the PV sum rules \eqref{sum-rules-1} would hold in Minkowski spacetime also for interacting fields to first order in perturbation theory. In addition it also been shown that for the models with exact supersymmetry all vacuum ultraviolet divergences are totally cancelled} for those (at least) finite or not finite quantum field theories that have $\gamma_n=0$ or all $\gamma_n=\gamma$.

 It is worth to mentioned the fact that in this work we concentrated only in Minkowski spacetime and did not consider the connection of the zero point energy density with the cosmological constant problem. However, even at this stage there are few comments in order. A possible extension of the PV sum rules in the case of de Sitter space has already been proposed in Ref. \cite{Kamenshchik:2006nm} in the case of free fields. It has been shown there that only the second and third on-shell PV sum rules get modified and turn out to be proportional to the Hubble parameter $H$. The extension of these sum rules in other curved spacetimes first require an adequate formulation of the problem in curved spacetime and second is beyond the main scope of the present paper. In connection with the cosmological constant, our conclusions only confirm what have already been presented in Refs. \cite{Zeldovich:1968ehl} - \cite{Koksma:2011cq} and \cite{Visser:2016mtr} (see also Ref. \cite{Martin:2012bt} for a review), namely that the zero point energy density does not scale like $\Lambda^4$ as wrongly assumed in the literature but it scales logarithmically with $\Lambda$ as we have shown by using three different regularization methods in Sec. \ref{sec:7}.
Even in this case the energy density of the vacuum is still quite large with respect to the energy density associated to the effective cosmological constant. However, this energy density is much smaller than that usually claimed in the literature where $\rho\propto M_\text{pl}^4$ with $M_\text{pl}$ being the Planck mass. A possible reduction of the value of $\Lambda_\text{eff}$ has been proposed in Refs. \cite{Mannheim:2010hk} and Ref. \cite{tHooft:2011aa}, where the gravity sector has been introduced in addition to the particle physics sector.

It is important to say few words about the assumptions that have been used in order to derive the PV sum rules. In this work the PV  sum rules have been found by requiring that the zero point stress-energy tensor is Lorentz invariant. However, these sum rules can also be found by using the original Pauli's argument, namely that the net zero point energy density must vanishes. Obviously this requirement is quite \emph{ad hoc} and there is no reason a priori that the net zero point energy must vanish. On the other hand, the requirement of Lorentz invariance of the zero point stress-energy tensor is a common practice in quantum field theory. Consequently, this requirement seems to be less \emph{ad hoc} assumption than the requirement of cancellation of the net zero point energy density. At this point one might ask; is the requirement of Lorentz invariance of the zero point stress-energy tensor justifiable? Typically in quantum field theory one expects that the zero point energy and pressure associated with the vacuum state in Minkowski space to be independent of the reference system. In addition, if our current understanding of the Minkowski vacuum as a state where all particles have the minimum energy possible and which is isotropic and homogenous in space is correct, then the zero point stress-energy tensor must be Lorentz invariant.

 Last thing which is worth to mention is that all particles of the SM model have been considered as elementary particles. In fact at the current state of particle physics, it seems that all SM particles are elementary rather than composite. However, in the hypothesis that one or more particles of the current SM that are considered elementary, would turn to be composite particles, then the right hand side of the sum rules \eqref {BSM-sum-rules} needs to be corrected.

 \vspace{1.5cm}

 {\bf{AKNOWLEDGMENTS}}:
 This work is supported by the Grant of the President of the Russian Federation for the leading scientific Schools of the Russian Federation, NSh-9022-2016.2.

\end{document}